\documentclass[aps,prd,twocolumn, amssymb, amsmath, amsfonts, showpacs, floatfix, twocolumn, twoside, a4paper, superscriptaddress, longbibliography, nofootinbib]{revtex4-2}
\usepackage{graphicx}
\usepackage{xcolor}
\usepackage{bm}
\usepackage{cancel}
\usepackage{comment}
\usepackage{makecell}

\usepackage[
    colorlinks=true,
    linkcolor=blue,
    citecolor=blue,
    urlcolor=blue
]{hyperref}

\allowdisplaybreaks[4]
\begin{document}

\title{Probing ultralight bosons with LISA observations of spinning black hole mergers and follow-up searches of merger remnants}
\author{Ifigeneia Giannakoudi}
\email{igiannakoudi@perimeterinstitute.ca}
\affiliation{Perimeter Institute for Theoretical Physics, Waterloo, Ontario N2L 2Y5, Canada}
\affiliation{Department of Physics \& Astronomy, University of Waterloo, Waterloo, Ontario N2L 3G1, Canada}
\author{Maxence Corman}
\email{maxence.corman@aei.mpg.de}
\affiliation{Max Planck Institute for Gravitational Physics (Albert Einstein Institute), D-14476 Potsdam, Germany}
\author{William E. East}
\email{weast@perimeterinstitute.ca}
\affiliation{Perimeter Institute for Theoretical Physics, Waterloo, Ontario N2L 2Y5, Canada}

\date{\today}

\begin{abstract}
    Ultralight bosons can trigger superradiant instabilities around 
    rotating black holes, extracting angular momentum and leading in some cases to observable gravitational signatures.
    When the associated spin-down timescale is shorter than the black hole lifetimes and the spin-up timescales due to, e.g. accretion, this process imposes an upper limit on black hole spins. In addition, the formation and subsequent dissipation of boson clouds generates quasi-continuous gravitational wave emission. In this work, we explore the prospects for constraining and detecting ultralight bosons with observations of massive black hole binary mergers with the space-based LISA observatory. We consider two complementary approaches: measurements of black hole spins from merging binaries and follow-up gravitational wave searches targeting massive black hole binary merger remnants. We consider three population models for massive black holes, based on either heavy or light seeds, 
    and forecast the exclusion and detection probabilities for both scalar and vector bosons.
    We find that black hole spin measurements can constrain scalar masses in the range $[5\times10^{-18},10^{-14}]$ eV and vector masses in the range $[6\times10^{-19},2\times10^{-14}]$ eV, with the exact range depending on the model. 
    In contrast, restricting to vector bosons, follow-up gravitational wave searches are sensitive to a narrower vector boson mass range of  $\sim[3\times10^{-17},3\times10^{-15}]$ eV, with the specific values again depending on the model. If a vector boson with a mass in the range $[10^{-16},2\times10^{-15}]$ eV existed, the probability of having an event that the follow-up searches would be sensitive to ranges from very small to near unity depending on the astrophysical model.
\end{abstract}

\maketitle

\section{Introduction}\label{sec.Intro}
Gravitational wave and black hole observations provide a powerful avenue to probe physics beyond the Standard Model. A well-studied example of this is the unique opportunity to constrain or potentially detect ultralight bosons indirectly, through black hole superradiance \cite{Starobinskii:1973vzb,Cromb:2020ldn}.
Such ultralight bosons arise naturally in a wide range of extensions of the Standard Model and general relativity. Examples include the QCD axion \cite{Kim_2016}, Majorons associated with lepton number symmetry breaking \cite{Chikashige:1980ui}, and ultralight particles emerging from string theory moduli compactifications~\cite{Arvanitaki_2010}. Ultralight vector bosons are also well motivated, with dark photons providing a prominent example \cite{Holdom1986,Ackerman_2009,Nakayama_2019}. In addition, massive spin-2 fields can arise in modified theories of gravity and have been proposed as a possible explanation for the cosmic acceleration \cite{Clifton:2011jh,Hinterbichler_2012,de_Rham_2014}.

Superradiant instabilities provide a mechanism through which rotating black holes can probe the existence of ultralight bosons and are most effective for particles whose gravitational coupling satisfies $\frac{GM \mu}{\hbar c^3} \lesssim 1$, where $M$ is the black hole mass and $\mu$ the boson rest energy. When bosonic modes satisfy the superradiant condition, waves scattering off a rotating black hole are amplified. For massive fields, the boson mass means it can be gravitationally bound, allowing the amplified field to grow into a bosonic cloud at the expense of the energy and angular momentum of the black hole. The instability ceases once the superradiant condition is saturated, leaving behind a black hole with reduced mass and spin, while the bosonic cloud subsequently dissipates through the emission of gravitational waves. A more detailed discussion of the superradiant instability is given in Sec.~\ref{sec.Theory}.

This process provides an observational window to detect or constrain ultralight bosons with both ground-based and space-based gravitational-wave detectors. Ground-based interferometers, operating in the $\sim 10$ to $10^3$ Hz band, are sensitive to stellar-mass black holes ($\sim 1$ to $100\,M_\odot$) \cite{2015,Acernese_2014,Aso_2013,akutsu2020overviewkagradetectordesign}.
The Laser Interferometer Space Antenna (LISA), with sensitivity in the $\sim 10^{-4}$ to $1$ Hz frequency range, will probe, among other systems, mergers of massive black holes with masses $10^4$ to $10^7\,M_\odot$~\cite{amaroseoane2017laserinterferometerspaceantenna}. Taken together, ground- and space-based observations probe black holes spanning roughly seven orders of magnitude in mass, from $\sim 1$ to $10^7\,M_\odot$. Since superradiance is most efficient when the boson Compton wavelength is comparable to the black hole radius, LISA is particularly sensitive to boson masses in the range $\sim 10^{-18}$ to $10^{-10}$ eV.

Gravitational wave observations can probe ultralight bosons through several complementary channels. Current searches include measurements of black hole spins in compact binary mergers \cite{Arvanitaki_2015,Brito:2017zvb,Cardoso_2018,Ng2021,caputo2025,GWTC-4_intro,Baryakhtar2017,aswathi2025}, searches for a stochastic gravitational wave background generated by boson clouds \cite{Tsukada2019,Yuan2022,Tsukada2021}, searches for continuous gravitational waves emitted by boson clouds \cite{O3_all-sky_scalar_bosons,Palomba2019,O3_galactic_center,Zhu2020,Dergachev2019}, and targeted follow-up searches for signals from post-merger remnants \cite{LIGOScientific:2025csr,Ghosh:2018gaw,Jones:2024fpg,Chan:2022dkt}. 
Spin measurements are particularly attractive because they rely on the robust prediction that superradiance extracts angular momentum from rapidly rotating black holes. 
Current constraints from ground-based detectors are often limited by large  uncertainties in the inferred spins, but there have been several recent observations of events with strong evidence for significant black hole spin
~\cite{LIGOScientific:2021usb,LIGOScientific:2025rsn,LIGOScientific:2025brd}, 
which have been used to place constraints on ultralight bosons using spin measurements \cite{Ng_2021,Ng:2020jqd,Ng:2019jsx,aswathi2025,Caputo:2025oap,LIGOScientific:2025brd,Ning:2026ebu}.
Additional constraints have been obtained from electromagnetic measurements of stellar mass and massive black hole spins \cite{Mummery:2025tvq,Reynolds:2019uxi,McClintock:2013vwa}. While these can probe a range of boson masses, they rely on astrophysical modeling and assumptions regarding black hole accretion, and may be less robust in scenarios where the boson couples weakly to the Standard Model.

Searches for a stochastic background and continuous waves are complementary, but their interpretation generally depends on assumptions about the properties of the underlying black hole population about which there is significant uncertainty.
Follow-up gravitational wave searches offer a complementary and comparatively clean probe. In this case, the properties of the remnant black hole are inferred directly from the observed merger signal, substantially reducing astrophysical uncertainties.
Existing follow-up searches have targeted remnants of LIGO-Virgo-KAGRA events \cite{LIGOScientific:2025csr}. Extending these searches to LISA sources in the future will provide sensitivity to lower boson masses, bridging the gap between ground-based gravitational wave constraints and those derived from electromagnetic spin measurements.

LISA observations are particularly promising for two reasons. First, massive black holes are expected to possess larger spins on average due to accretion, increasing the efficiency of superradiant growth \cite{Thorne:1974ve,Barausse:2012fy, Barausse_2023}. Second, LISA is expected to observe a large population of high signal-to-noise ratio massive black hole mergers, enabling precise measurements of black hole masses and spins \cite{Barausse_2023,colpi2024lisadefinitionstudyreport}.
Consequently, both spin measurements and follow-up searches with LISA have the potential to place powerful constraints on ultralight bosons. In the absence of detectable boson-cloud signals, follow-up searches can exclude regions of boson parameter space, while black hole spin measurements can rule out masses for which superradiance would have efficiently spun down the observed systems \cite{Isi_2019,chan2022extractingultralightbosonproperties,Jones_2023,Arvanitaki_2015,Arvanitaki_2017,Cardoso_2018,Ng2021,Baryakhtar2017,stott2020ultralightbosonicfieldmass,aswathi2025}.
Realizing the full potential of these approaches, however, requires accurate modeling of both the binary merger signals and the post-merger gravitational wave emission from boson clouds.

In this paper, we estimate the range of ultralight scalar and vector boson masses that can be constrained or detected 
with LISA 
through spin measurements and follow-up searches, and quantify the corresponding exclusion and detection prospects.
We make use of \texttt{SuperRad}\footnote{\url{http://www.bitbucket.org/weast/superrad}}, an open source black hole 
superradiance waveform model incorporating
theoretical predictions for black hole spin-down 
and gravitational wave observables 
in the relativistic and non-relativistic regimes
~\cite{Siemonsen:2022yyf,May:2024npn}.
Our analysis is based on three massive black hole population models \cite{Barausse:2012fy}, using catalogs introduced in Ref.~\cite{Klein:2015hvg} and updated in Ref.~\cite{Barausse_2023} following recent pulsar timing array observations~\cite{antoniadis2024seconddatareleaseeuropean,Afzal_2023,Figueroa_2024}, which may be interpreted as evidence for a stochastic background from massive black hole binaries \cite{Agazie_2023_BH}.

Using the most recent catalogs for these three massive black hole population models, we estimate the constraints on ultralight boson masses that can be obtained with LISA through both black hole spin measurements and targeted post-merger follow-up searches. We find that spin measurements can exclude scalar and vector boson masses in the range $\sim[10^{-18}-10^{-14}]$ eV, with specific values depending on the boson type and black hole population model. Targeted follow-up searches are estimated to constrain vector boson masses in the range $[10^{-17}-10^{-15}]$ eV.  In the event that ultralight vector bosons exist, we find that a narrower mass window around $10^{-16}$ eV may give detectable post-merger signals, although the prospects depend on the underlying black hole population.

Previous studies have investigated the prospects for constraining ultralight bosons with LISA using black hole spin measurements, all-sky searches for resolvable monochromatic gravitational wave signal, and measurements of the stochastic gravitational wave background~\cite{Brito:2017zvb,Brito:2017wnc}. Our scalar boson results are broadly consistent with these earlier analyses. This paper extends them in three directions. First,
we include both scalar and vector bosons. Second, we employ massive black hole population catalogs that have been updated in light of recent observations~\cite{Barausse:2012fy,Barausse_2023}. 
Third, we investigate targeted post-merger follow-up searches as a complementary observational strategy, expanding on the calculation of the detectability of boson cloud signals from individual merger remnants done in Ref.~\cite{Siemonsen:2022yyf}, to consider a population based event rate forecast while consistently accounting for the effects of superradiant evolution on the progenitor black holes. These improvements provide updated and complementary projections for the ability of LISA to probe ultralight bosons across a broad range of masses.

The remainder of this paper is organized as follows. In Sec.~\ref{sec.Theory}, we review the theory of 
black hole superradiance and bosonic instabilities. In Sec.~\ref{sec.Method}, we describe our methodology 
for estimating exclusion and detection prospects. In Sec.~\ref{sec.Results}, we present and discuss our results. We conclude in Sec.~\ref{sec.Discussion}.

\section{Theory}\label{sec.Theory}
\subsection{Black Hole Superradiant Instabilities from Ultralight Bosons}
In this section, we briefly review black hole superradiance and the associated instabilities relevant for ultralight bosons. We use units with $\hbar=c=G=1$.

Superradiance refers to the enhancement of radiation when scattering off a system from which energy can be extracted in the presence of dissipation. Rotating black holes, described by the Kerr solution \cite{Kerr:1963ud}, provide a natural example of such a system \cite{Zeldovich1971,Zeldovich1972,Bekenstein:1973ur,Vilenkin:1978uc}. A key feature of the Kerr spacetime is the existence of an ergoregion, bounded by the ergosurface and the event horizon, within which no static observers can exist. In this region, the Killing vector associated with stationarity becomes spacelike, allowing for timelike particles to have negative Killing energies and enabling energy extraction from the black hole \cite{Vicente_2018}. 

A bosonic field of real frequency $\omega_R$ and azimuthal number $m$ incident on a spinning black hole will induce a flux of energy $E$ and angular momentum $L$ across the black hole horizon with ratio $L/E = m/\omega_R$.
Fields satisfying the superradiance condition 
\begin{equation}\label{eq.BH superradiance condition}
    \omega_R < m \Omega_H ,
\end{equation}
where $\Omega_H$ is the black hole horizon frequency
extract energy and angular momentum from the black hole.

For massless fields, amplified modes escape to infinity and no instability develops. However, for bosonic fields with non-vanishing rest mass, the mass term provides an effective confining potential, allowing for the formation of states that are gravitationally bound to the black hole and have a continuous flux across the horizon. When these states satisfy the superradiance condition, repeated amplification leads to an instability and exponential growth of a bosonic cloud around the black hole. Although the above assumes an adiabatic description \cite{Brito_2015}, this has been found to be a good approximation even in the fully non-linear scenario \cite{East_2018,East:2017ovw}. The black hole loses mass and angular momentum until it spins down to the point where the superradiance condition is approximately saturated. During the growth of the cloud, the field can extract up to  $ \sim 10 \%$ of the black hole's mass until the instability is saturated and the black hole cloud configuration reaches a quasi-equilibrium configuration. The cloud then dissipates through gravitational-wave emission. The growth and emission timescales are typically well separated, hence these phases can be treated independently. Since higher azimuthal modes will still satisfy the superradiance condition, they can continue to grow, though on longer timescales.

A key parameter controlling this process is the gravitational fine structure constant, 
\begin{equation}\label{eq.alpha}
    \alpha = \frac{r_g}{\lambda_C} = M \mu \, ,
\end{equation}
which measures the ratio of the black hole gravitational radius to the boson Compton wavelength. For $\alpha \ll 1$, the cloud is located farther away from the black hole, the instability develops slowly and can be treated non-relativistically. For $\alpha \sim \mathcal{O}(1)$, the cloud sits close to the black hole, the growth rate is significantly enhanced, and relativistic effects are therefore important.

Superradiant instabilities have been extensively studied for scalar (spin-0), vector (spin-1), and tensor (spin-2) fields. In this work, we focus on scalar and vector bosons as the backreaction of massive spin-2 fields is complicated and potentially model dependent~\cite{May:2025arz}. Their respective dynamics on a Kerr background are governed by
\begin{equation}\label{fieldeqs}
(\Box - \mu_S^2)\Phi = 0, \quad \nabla_\mu F^{\mu\nu} = \mu_V^2 A^\nu,
\end{equation}
where $\Box\equiv \nabla_\mu\nabla^\mu$ is the covariant wave operator and $F_{\mu\nu}\equiv \nabla_\mu A_\nu-\nabla_\nu A_\mu$ is the vector's field strength tensor.
Due to the axisymmetry and stationarity of the Kerr spacetime, solutions to the above field equations can be
written in the form
\begin{equation}\label{fieldansatz}
\Phi, A_\mu \sim e^{-i(\omega t - m \varphi)} ,
\end{equation}
where we introduced the complex frequency $\omega = \omega_R + i \omega_I$ which encodes
both the oscillation frequency of the cloud which is half of the characteristic gravitational wave frequency $f_{\rm GW}= \omega_R/ \pi$, and the instability growth
timescale $\tau_I = 1/\omega_I$ (where we take the convention that $\omega_I > 0$ for a growing mode). 

The cloud emits quasi-monochromatic gravitational waves. The gravitational wave strain can be written as
\begin{equation}\label{hsff}
h = h_+ - i h_\times = \frac{\mathcal{A}}{r} e^{-i \phi_{\rm GW}(t)} \psi(\theta) e^{i m_{\rm GW} \varphi},
\end{equation}
and the gravitational wave frequency is twice the cloud oscillation frequency, hence
\begin{equation}
    \phi_{\rm GW}(t) = 2 \int \omega_R (t) dt.
\end{equation}
The azimuthal dependence is fixed by that of the cloud with $|m_{\rm GW}| = 2 m$, whereas the polar contribution $\psi(\theta)$ is dominated by the $\ell_{\rm GW}=|m_{\rm GW}|$ spin-(-2) weighted spherical harmonic mode. 

In this work, we use the \texttt{SuperRad} \cite{Siemonsen:2022yyf,May:2024npn} package to model the evolution of the boson cloud and extract the relevant physical quantities governing black hole spin-down and gravitational wave emission. 

\subsection{Black Hole Population Models}\label{theory:models}
In order to estimate the constraints that can be placed on ultralight bosons with LISA, we need a population of merging massive black hole binaries. There are still large uncertainties regarding the populations of such binaries that LISA will observe \cite{PhysRevD.83.044036,Barausse:2020mdt}. Therefore, we have to rely on simulations and consider several different possible scenarios for how black holes form and grow \cite{Volonteri:2002vz,Sesana:2007sh,Plowman:2010fc,Rodriguez-Gomez:2015aua,Ricarte_2018,Bonetti:2018tpf,Dayal_2019,Barausse:2020mdt,Klein:2015hvg,PhysRevD.83.044036}. Semi-analytical models have been established as one of the possible methods to predict the population of merging massive black hole binaries. In this work, we adopt the semi-analytical model developed
in Ref.~\cite{Barausse:2012fy} (with contributions from Refs.~\cite{Klein:2015hvg,Antonini:2015sza,Barausse:2020mdt,Antonini:2015cqa,Sesana:2014bea} and more recently Ref.~\cite{Barausse_2023}) to track the evolution of massive black holes across cosmic time. We consider three different astrophysical models with different seed and time-delay prescriptions:
\begin{itemize}
  \item \textbf{PopIII Model}: A light-seed scenario where massive black holes are assumed to form at high redshift ($z > 15-20$) in the mass range $10^2-10^3\ M_{\odot}$ from the collapse of
  heavy metal-poor population III stars. 
  This model takes into account the delay between the host galaxy merger and the massive black hole binary merger.
  \item \textbf{Q3 delays Model}: A heavy-seed scenario in which most of the mass in the protogalactic disk collapses into a supermassive star leaving behind a black hole in the mass range
  $10^4-10^5\ M_{\odot}$ at redshifts $z \sim 8-15$.  This model also accounts for the delay between galactic and massive black hole binary mergers.
  \item \textbf{Q3 no delays Model}: A similar heavy-seed scenario but without delays between the
  galaxy and black hole merger leading to more events (at higher redshift). This scenario can be considered as optimistic in the number of predicted massive black hole binary mergers.
\end{itemize}

These models predict the merger rate, the intrinsic binary properties 
(masses, spin magnitudes and orientations, and luminosity distances) and the properties of 
the host galaxy (amount of mass in gas and stars, mass in the disk, etc.). We refer the reader to the aforementioned papers \cite{Barausse:2012fy,Barausse_2023} for more details. 
As previously concluded in Refs.~\cite{Barausse_2023,antoniadis2024seconddatareleaseeuropean,Afzal_2023,Agazie_2023_BH}, recent
pulsar timing observations \cite{PTA_2023, Tarafdar_2022, PTA_Agazie_2023, Reardon_2023, Xu_2023} seem to disfavor large delays at separations of hundreds of parsecs and instead indicate massive black hole binaries merge efficiently after their host galaxies merge. Therefore, we only consider models with no delay or medium time delays in agreement with pulsar timing array data. Models at finite resolution, as well as those extrapolated to infinite resolution are publicly available\footnote{\url{https://people.sissa.it/~barausse/catalogs/}}. Here we use the infinite resolution simulations whose theoretical predictions agree with pulsar timing measurements~\cite{Barausse_2023}.

\subsection{Spin-down Timescale}
A key ingredient in our analysis is the timescale over which superradiance can extract angular momentum from a black hole. In order for superradiance to leave an observable signature on the black hole spin distribution, 
the instability must spin down the black hole faster than astrophysical processes that increase its 
angular momentum. For supermassive black holes, the dominant competing process is accretion. Since these
objects reside in galactic centers and are typically surrounded by gas and plasma, accretion can efficiently increase both their mass and spin. The characteristic mass growth and spin up timescales are comparable
and can be approximated by the accretion timescale
\begin{equation}
\tau_{\rm acc} \sim \frac{T_S}{f_{\rm Edd}}  , 
\end{equation}
where $T_S$ is the Salpeter time, $f_{\rm Edd} \equiv \dot M / \dot{M}_{\rm Edd}$ is the Eddington ratio, and we assume
a radiative efficiency of $\eta \sim 0.1$ \cite{Brito_2015}.
For Eddington limited accretion [$f_{\rm Edd}\sim \mathcal{O}(1)$], this corresponds to a characteristic timescale of approximately
\begin{equation}
    T_S \simeq 4.5\times10^7 \ {\rm yr},
\end{equation}
which we adopt as our fiducial upper limit for the superradiant spin-down timescale. Therefore, if the instability develops on longer timescales, accretion is expected to dominate and erase the spin-down signature.
In practice, accretion histories are highly uncertain and depend on the properties of the galactic environment \cite{2010A&ARv..18..279V,2011MNRAS.417.2085V,Kormendy:2013dxa,Harrison:2024cpw}. Sub-Eddington systems may experience feedback through jets \cite{Yuan:2014gma,2012MNRAS.420.2662D}, while near-Eddington and super-Eddington systems can be regulated by winds, radiation pressure and episodes of accretion \cite{1973A&A....24..337S,Dotan:2010wu,Giustini:2019she,Proga:2000bw,Inayoshi:2015pox}. Simulations of super-Eddington 
accretion suggest that feedback from powerful jets can regulate accretion on timescales of a few million years \cite{Massonneau:2022uwg}. 

 Another mechanism that can affect black hole spin evolution is the Blandford–Znajek mechanism \cite{1977MNRAS.179..433B}, in which rotational energy is extracted electromagnetically through jets.
 In magnetically arrested disk states~\cite{1974Ap&SS..28...45B,2003ApJ...592.1042I,2003PASJ...55L..69N}, this process can reduce black hole spins, with characteristic 
 timescales ranging from $10^6 - 10^9$ years depending on the accretion rate and magnetic flux  \cite{Ricarte:2023owr,Narayan:2021qfw}. Since the efficiency of this process decreases as the black hole spins down, adopting the Salpeter timescale remains a reasonable first-order approximation. To further test the robustness of our results given the uncertainties, we additionally consider an even more conservative timescale of 
$\tau_{\rm sd} = 0.01 T_S$.
 
Another timescale of relevance is that of massive black hole mergers. Supermassive black hole growth is thought to proceed through a combination of hierarchical mergers and accretion \cite{2003ApJ...582..559V,2005Natur.433..604D,2009MNRAS.400.1911V}. 
Merger timescales depend strongly on the galactic environment and mass ratio, ranging from a few Myr to 
several Gyr \cite{Fang:2022cso,Sesana:2006ne}. A companion star or black hole may perturb or disrupt a bosonic cloud, particularly if the cloud is diffuse and extends far from the horizon \cite{Baumann:2018vus,Baryakhtar_2021}. While such effects, as well as level
mixing induced by gravitational perturbations, may be relevant ~\cite{Baumann_2020,Zhang_2020,Zhang_2019,Baumann:2018vus}, they are beyond the scope of this work. 

Throughout this work, we assume isolated black hole-cloud systems and purely gravitational interactions. Our adopted spin-down timescale thresholds are shorter than typical merger timescales and therefore provide a consistent framework for estimating superradiant spin depletion.

\section{Methods}\label{sec.Method}
In this section, we describe the methods used to assess the prospects for probing ultralight bosons with LISA observations. We consider two complementary approaches: constraints derived from black hole spin measurements inferred from the binary merger gravitational wave signal, and targeted follow-up searches for gravitational wave signals from the merger remnant. We also examine the scenario where an ultralight vector boson exists, and therefore modifies the allowed spins of binary black holes, and estimate the prospects for detecting the resulting boson cloud through post-merger follow-up observations. We begin by describing how we construct the synthetic massive black hole binary catalogs used in this analysis.

\subsection{Catalog construction}
The massive black hole binary catalogs introduced in Sec.~\ref{theory:models} provide the redshift, the component masses and spins, and a weight proportional to the host halo number density. We convert these weights into merger rates by multiplying by the corresponding comoving volume element.  
Assuming an observing time of $4$ years, each catalog entry is treated as an independent Poisson process with mean given by this rate. Synthetic populations are generated by drawing the number of occurrences of each system from the corresponding Poisson distribution. We generate $5000$ realizations for the Q3delays and Q3nodelays models and $3100$ realizations for PopIII~\footnote{The number of realizations for each catalog was chosen so that the mean number of mergers with signal-to-noise ratio (SNR) above $8$ converged to the numbers reported in Table I of Ref.~\cite{Barausse_2023} with standard error of order $0.1$. For PopIII, which has the largest catalog,  $3100$ realizations were sufficient.}. A merger time, uniformly distributed between one week and four years is assigned to each event. For each synthetic catalog, we compute the expected gravitational wave signal and retain only events that satisfy our detectability criterion described below.

In each realization, we model the gravitational wave signals of the massive black hole binaries using the \texttt{lisabeta} code described in Ref.~\cite{Marsat:2020rtl}. We make use of the IMRPhenomHM~\cite{London:2017bcn} waveform model, which is a non-precessing model that describes binaries with aligned spins, but which includes contributions from higher order modes.
We then keep only the events whose sky, inclination, and polarization averaged gravitational wave SNR is above 20\footnote{In the literature, the standard observational threshold is usually an SNR above 8. We select 20 as our threshold 
because we are interested in events in which the black hole spins can be well measured or in which one could measure a (weaker) signal from the post-merger remnant.}. The SNR is computed according to
\begin{equation}\label{SNR}
\mathrm{SNR}^2 = 4 \int_{f_{\min}}^{f_{\max}} \frac{\left<|\tilde{h}(f)|^2\right>_{\theta}}{S_n(f)} \, df , 
\end{equation}
where $\tilde{h}(f)$ corresponds to the Fourier transform of the time-domain signal, the brackets denote sky position and orientation average, $[f_{\min}, f_{\max}]$ are the frequency bounds corresponding to the duration of the mission, which are limited to the range $[10^{-5},0.5]$ Hz, and $S_n(f)$ is the noise power spectral density. 
Following Ref.~\cite{Robson_2019}, we assume the total noise spectral density is given by $S_n(f) = P_n(f)/ \mathcal{R}(f) + S_c(f)$, where $P_n(f)$ is the power spectral density of the detector noise, $\mathcal{R}(f)$ is the sky and polarization averaged signal detector's response function and $S_c(f)$ is an estimate of the galactic confusion noise. For simplicity, in the inclination averaging calculation we include only the dominant $\ell=|m|=2$ mode. While higher order modes can contribute significantly to the signal and improve parameter estimation for some massive black hole binaries, particularly at high masses near merger~\cite{Pitte2023, Yi2025, Yi2026}, 
our analysis is restricted to estimating detectability through a sky-, polarization-, and inclination-averaged SNR. Restricting the calculation to the quadrupolar mode provides a conservative estimate of the detection prospects.

\begin{figure*}[htbp] 
    \centering
    \includegraphics[width=0.4\textwidth]{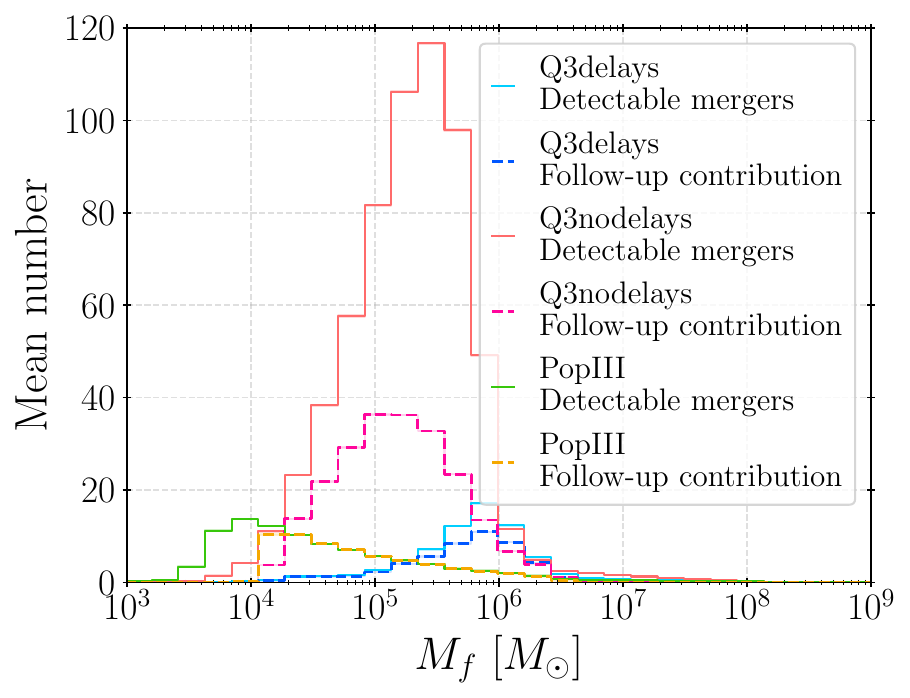}
    \includegraphics[width=0.4\textwidth]{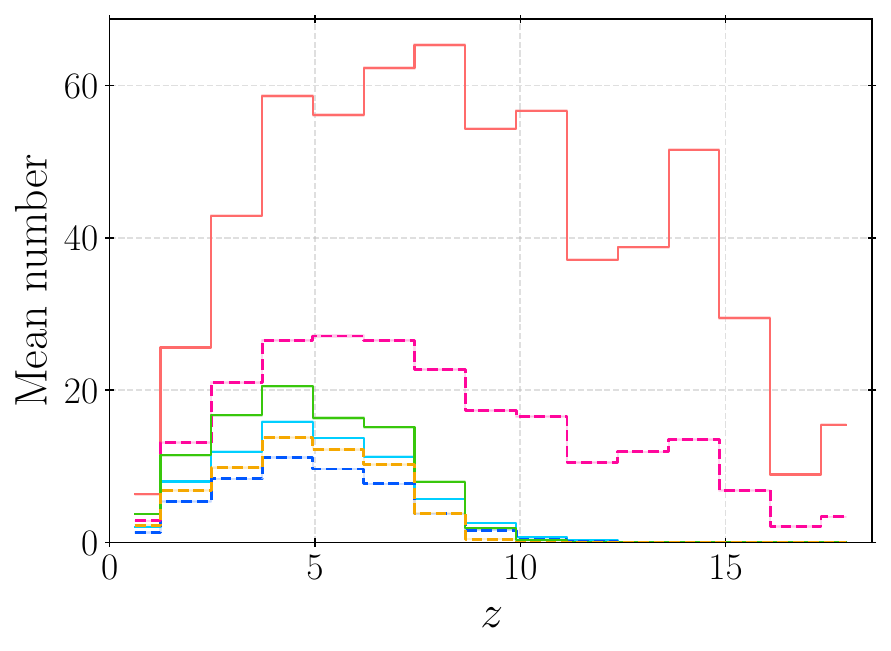}
    \caption{Left panel: Mean number of detectable binaries (SNR $\geq20$) that merge during 4 years, binned according to remnant mass, as well as the mean number that contribute to superradiance follow-up signals (SNR $\geq10$). Right panel: the same mean number of mergers binned according to redshift. We include three black hole population models: Q3delays, Q3nodelays, and PopIII. }
    \label{fig.1d_distr}
\end{figure*}

Figure~\ref{fig.1d_distr} shows the mean number of detectable mergers as a function of remnant mass (left panel) and redshift (right panel) for the three astrophysical models considered in this work. The PopIII light-seed model (green) is shifted towards lower remnant masses than the heavy-seed Q3 models (coral and blue), reflecting the different seed formation scenarios. In redshift, all models span a broad range, although the Q3delays model is preferentially concentrated at lower redshifts because of the delays between galaxy and black hole mergers included in that model. While PopIII seeds form at earlier cosmic times, the model's detectable merger population does not extend to substantially higher redshifts than the Q3nodelays model. This is because the lower black hole masses characteristic of PopIII mergers produce weaker signals, causing a larger fraction of high redshift events to fall below the detectability threshold.

Most detectable mergers have remnant masses in the range $10^4-10^7$ \(M_\odot\) and occur at redshifts $z \lesssim 10$, corresponding to the region where LISA is expected to be the most sensitive to massive black hole mergers. Assuming a four year mission, the Q3delays, Q3nodelays, and PopIII models predict on average 73, 647 and 250 detectable mergers, respectively. These numbers are in close agreement with those reported in Table I of Ref.~\cite{Barausse_2023}, despite differences in the detectability criterion and waveform treatment. 

Since the efficiency of superradiance depends strongly on the black hole spin, it is also important to characterize the spin distribution of the detectable population. The left panel of Fig.~\ref{fig.2d_distr} depicts the mean number of detections within four years as a function of primary mass and dimensionless spin.
The three models exhibit distinct distributions. The PopIII model is concentrated at masses $10^4$--$10^5\ M_\odot$ with nearly extremal spins, while Q3delays peaks around $10^6\ M_\odot$ and dimensionless spins above $0.9$.
The Q3nodelays model exhibits a broader distribution, with primary masses concentrated around $10^5$--$10^6\,M_\odot$ and spins typically above $\sim0.7$. 
Several features of the mass-spin distributions can be understood from the spin evolution in the underlying population models. At moderate and high redshifts ($z\gtrsim4$-$5$), black holes typically reside in gas-rich environments where accretion is coherent, efficiently spinning them up toward extremal values.
Consequently, detectable black holes in the mass range $\sim10^6$ to $4\times10^6\ M_\odot$ in the Q3 models, and $\lesssim10^7\ M_\odot$ in PopIII, are found predominantly at very high spins.
At larger masses, however, the gas reservoir is progressively depleted and accretion becomes less coherent. This suppresses further spin-up and leads to the broader distribution of spins observed for the most massive black holes. These trends are consistent with the spin evolution found in the population synthesis models of Ref.~\cite{Barausse:2012fy}.

Next we describe the three different observational scenarios we consider for probing ultralight bosons.

\subsection{Exclusion From Spin Measurements}
 For sufficiently loud massive binary black hole merger signals observed by LISA, the component masses and spins can be measured with high precision. 
Parameter estimation studies indicate that spin uncertainties are typically at the level of 0.1\%-1\%~\cite{colpi2024lisadefinitionstudyreport}, and Fig. 6 of \cite{Barausse_2023} shows that approximately $40-60\%$ of detectable sources have spin measurement accuracies better than $1\%$, depending on the underlying population model. In our analysis, we adopt a conservative relative spin uncertainty of 10\%, and also consider a $1\%$ uncertainty for comparison, finding qualitatively similar results.
This treatment is approximate, and one could improve on it by inferring the distribution of recovered
spins for each signal realization. However, given the significant uncertainties associated with the underlying black hole population models, a simplified uncertainty description is sufficient for the purposes of this study.

For a black hole of mass $M$, a boson of mass $\mu$, and a spin-down timescale $\tau_{\rm sd}$, we compute the maximum spin this black hole can support, $a_{\rm max}(M,\mu,\tau_{\rm sd})$. If the measured spin of a black hole exceeds $a_{\rm max}$, after accounting for the assumed measurement uncertainty, the corresponding boson mass is considered excluded by that binary constituent. 

We consider ultralight boson masses in the range $\mu \in [10^{-23}, 10^{-9}] \, \text{eV} $ and two spin-down timescales, $T_S$ and $0.01 T_S$. Using the detectable binary populations described in the previous section, we evaluate $a_{\rm max}$ for every binary constituent in each realization. Constituents satisfying the exclusion criterion are counted as constraining the corresponding boson mass. We perform this procedure for both vector and scalar bosons. 

The probability of constraining a given boson mass with LISA is then defined as the fraction of realizations in which at least one constituent of at least one detectable binary excludes that mass. This quantifies the likelihood that LISA yields at least one exclusion event for a given boson mass during the mission lifetime. 

The computation of $a_{\rm max}$ follows Ref.~\cite{aswathi2025}. For a given black hole mass $M$, boson type and mass $\mu$, and spin-down timescale $\tau_{\rm sd}$, we use \texttt{SuperRad} \cite{Siemonsen:2022yyf,May:2024npn} to evolve the superradiant cloud for a range of initial black hole spins in the interval $[0,1)$, for which the superradiance condition is satisfied (if any) [Eq.~\eqref{eq.BH superradiance condition}]. The evolution tracks the growth of all unstable modes and updates the black hole mass and spin accordingly. A configuration is viable only if the total growth time of the instability (summed over all contributing modes) satisfies $\sum t_{c,m} \leq  \tau_{\rm sd}$. The maximum allowed initial spin for which evolution remains consistent with the condition defines $a_{\rm max}(M,\mu,\tau_{\rm sd})$, i.e. the largest spin a black hole can have while still being spun down by the boson cloud within a specified timescale.

For vector bosons, we use the relativistic \texttt{SuperRad} \cite{Siemonsen:2022yyf,May:2024npn} model for azimuthal numbers $m=1-5$, supplemented by the non-relativistic approximation for higher order modes. For scalar bosons, we use the relativistic treatment for $m=1$ and $2$ and the non-relativistic approximation for higher modes, up to $ m=15$. Tables \ref{table.vector} and \ref{table.scalar} illustrate the impact of a representative superradiant instability with boson mass $6 \times 10^{-17}$ eV on a black hole of mass  $M_\mathrm{BH} = 10^6 \,M_\odot$ and initial dimensionless spin $a = 0.995$. The results highlight the large hierarchy in growth timescales between the fastest growing mode and the higher modes, while also demonstrating that subdominant modes can collectively contribute non-negligibly to the spin-down. 
In this example, a nearly maximally spinning black hole is spun down to a spin of $0.472$ for the vector case and $0.481$ for the scalar case, within timescales shorter than the Salpeter time, $T_S \sim 4.5 \times 10^7 \,\mathrm{years}$.

\begin{table}[htbp]
\caption{Black hole mass and dimensionless spin after saturation of each superradiant mode with azimuthal number $m$, together with the corresponding cloud growth timescales and characteristic gravitational wave emission timescales, for a representative system with $M_\mathrm{BH} = 10^6\ M_\odot$ and initial spin $a = 0.995$, subject to superradiance instability induced by a vector boson of mass $\mu_V = 6 \times 10^{-17} \, \mathrm{eV}$. Modes with larger azimuthal number exhibit growth timescales exceeding the age of the Universe.}
\label{table.vector}
\renewcommand{\arraystretch}{1.25}
\begin{tabular}{c c c c c}
\hline
$m$ & $M_\mathrm{BH}\,[M_\odot]$ &  $a$ & $t_{c,m}$ [years] & $t_{\rm GW,m}$ [years] \\
\hline
1 & $9.49\times10^{5}$ & 0.963 & $7.598\times10^{-1}$ & $8.363\times10^{2}$ \\
2 & $8.69\times10^{5}$ & 0.666 & $4.003\times10^{2}$ & $9.742\times10^{7}$ \\
3 & $8.44\times10^{5}$ & 0.472 & $1.003\times10^{8}$ & $2.853\times10^{13}$ \\
\hline
\end{tabular}
\end{table}

\begin{table}[htbp]
\renewcommand{\arraystretch}{1.25}
\caption{Same as Table~\ref{table.vector} but for a scalar boson of the same mass.}
\label{table.scalar}
\begin{tabular}{c c c c c}
\hline
$m$ & $M_\mathrm{BH}\,[M_\odot]$ & $a$ & $t_{c,m}$ [years] & $t_{\rm GW,m}$ [years] \\
\hline
1 & $9.78\times10^{5}$ & 0.988 & $2.493\times10^{3}$ & $2.750\times10^{7}$ \\
2 & $8.88\times10^{5}$ & 0.683 & $1.535\times10^{5}$ & $1.112\times10^{14}$ \\
\hline
\end{tabular}
\end{table}

\subsection{Exclusion From Follow-up Searches}
Another approach to constraining ultralight bosons through black hole superradiance is via follow-up searches targeting merger remnants. In this scenario, a binary merger is first observed through gravitational waves, allowing for an estimate of the remnant mass and spin. If an ultralight boson exists with mass such that the superradiance condition [Eq.~\eqref{eq.BH superradiance condition}] is satisfied, a bosonic cloud will form around the remnant. This cloud subsequently dissipates through the emission of gravitational waves. 
If the cloud growth timescale of the fastest growing mode is shorter than the remaining mission duration, and if the resulting gravitational wave signal would be detectable, the non-observation of such a signal can be used to constrain the corresponding boson mass. We restrict this analysis to vector bosons, as scalar cloud growth timescales typically exceed the mission lifetime (this is illustrated in the fourth column of Tables \ref{table.vector} and \ref{table.scalar}).

For each remnant in our realizations with mass $M_f$ and spin $a_f$, we use \texttt{SuperRad} \cite{Siemonsen:2022yyf,May:2024npn} to determine
the vector boson cloud growth time, the gravitational wave emission timescale, the strain amplitudes of the two polarizations $(h_{+}, h_{\times})$, and the time-dependent frequency evolution $f_{\rm GW}(t)$. Using these quantities, we evaluate the sky/inclination/polarization averaged SNR defined in \eqref{SNR}. If the SNR is $\geq 10$ we classify the signal as detectable, and therefore its absence can be used to place constraints on the boson mass. 
We perform this procedure for boson masses in the range $\mu \in [10^{-23}, 10^{-9}] \, \text{eV} $.
The probability of constraining a given boson mass through follow-up searches is defined as the fraction of realizations containing at least one detectable post-merger boson cloud signal associated with that mass. Equivalently, this quantity represents the probability that LISA observes at least one remnant capable of excluding the corresponding boson mass during the mission lifetime.

Additional details on the computation of the sky/inclination/polarization averaged SNR starting from the time domain waveforms computed with \texttt{SuperRad} are provided in Appendix~\ref{app:SNR}.

\subsection{Detection Probability with Follow-up Searches}
In the previous subsections, we described how we could constrain boson masses through spin measurements and the absence of follow-up signals. Follow-up searches also offer discovery potential if an ultralight vector boson were to exist in the relevant mass range.
Unlike the exclusion analysis, which assumes that the observed binary properties are unaffected by the boson, the detection analysis must account for the impact of superradiant spin-down on the binary components prior to merger. As in the previous subsection, we restrict our analysis to vector bosons, since for scalars, boson clouds generally grow too slowly to produce detectable follow-up signals within the LISA mission lifetime under consideration.
For each vector boson mass considered, we apply the spin-down procedure described above to all binary components that satisfy the superradiance condition.
This modifies the component masses and spins at the time of merger.
We then recompute the properties of the merger remnants using Eqs. (54) and (57) of Ref.~\cite{Barausse_2023}, based on the fitting formulae of Refs.~\cite{Barausse:2009uz,PhysRevD.80.124026}. Finally, we repeat the analysis described in the previous subsection to estimate the number of systems that (i) merge during the LISA mission with SNR $\geq 20$, and (ii) can produce a follow-up signal with SNR $\geq 10$, after accounting for the impact of superradiant instability on the binary components. A boson mass satisfying these conditions is considered detectable with LISA through its associated post-merger boson cloud signal. The probability of detection for a given boson mass is defined as the fraction of realizations containing at least one detectable follow-up boson cloud signal associated with that mass.

\begin{figure*}[htbp]
\centering

\begin{tabular}{c c}

\textbf{Detectable primaries} & \textbf{Detectable primaries with vector $\mu\simeq 5\times 10^{-16}$ eV} \\[0.3em]

\makebox[0pt][r]{\raisebox{2.2cm}{\rotatebox{90}{Q3delays}}\hspace{0.5em}}%
\includegraphics[width=0.4\textwidth]{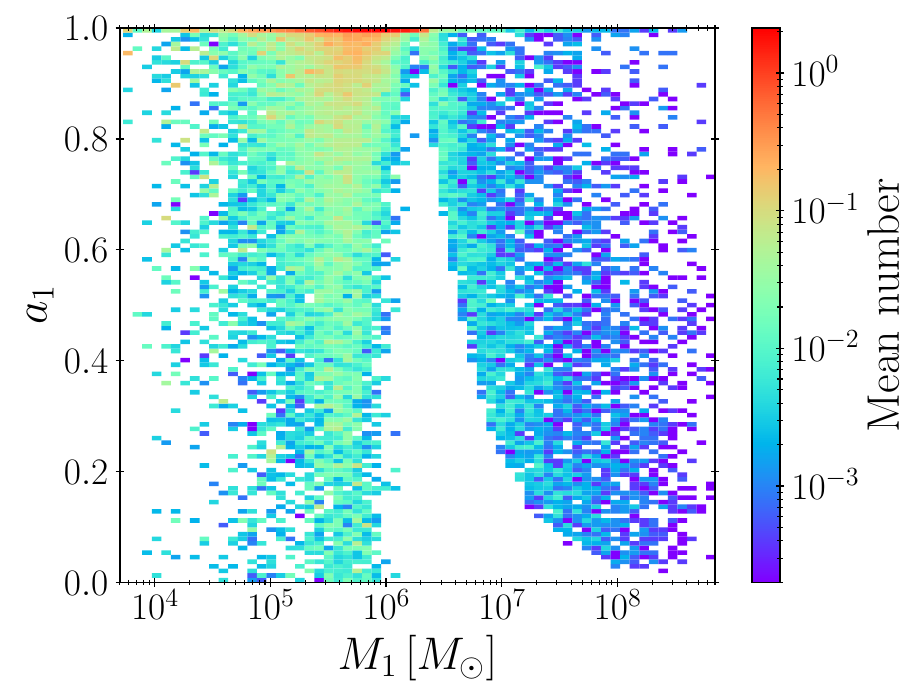}
&
\includegraphics[width=0.4\textwidth]{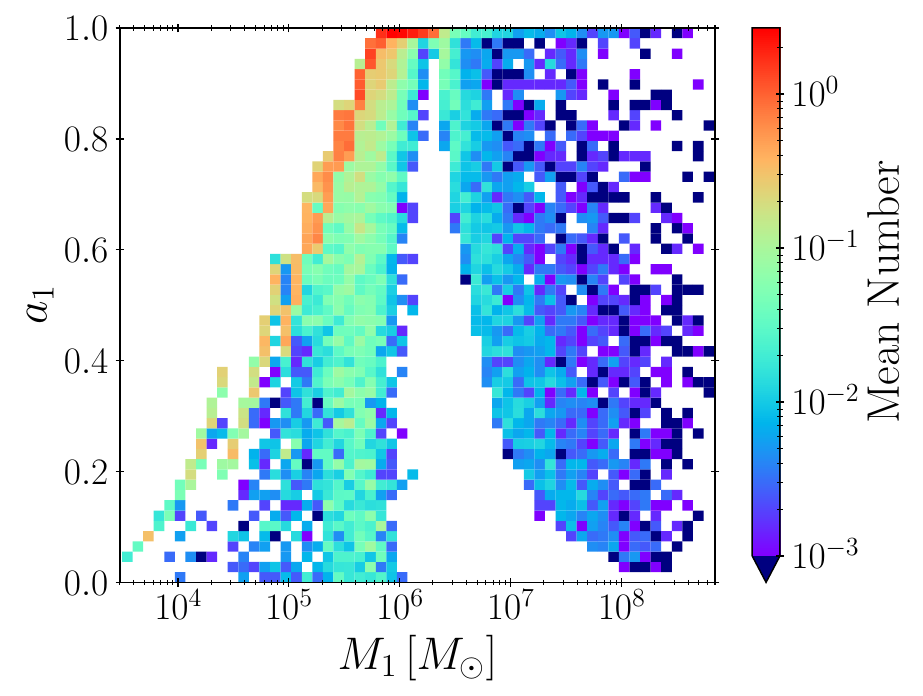}
\\[0.6em]

\makebox[0pt][r]{\raisebox{2.2cm}{\rotatebox{90}{Q3nodelays}}\hspace{0.5em}}%
\includegraphics[width=0.4\textwidth]{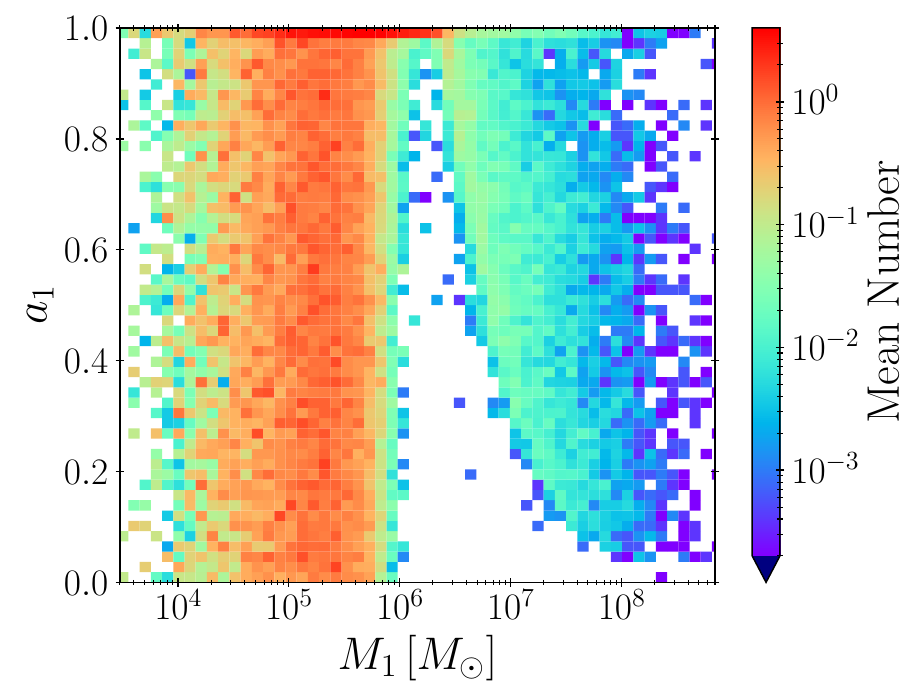}
&
\includegraphics[width=0.4\textwidth]{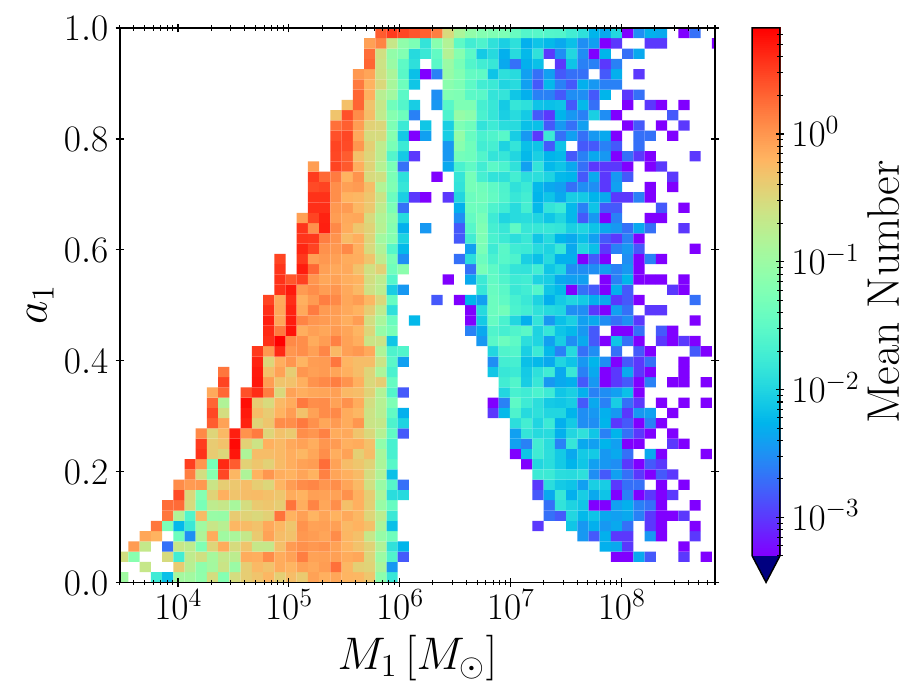}
\\[0.6em]

\makebox[0pt][r]{\raisebox{2.2cm}{\rotatebox{90}{PopIII}}\hspace{0.5em}}%
\includegraphics[width=0.4\textwidth]{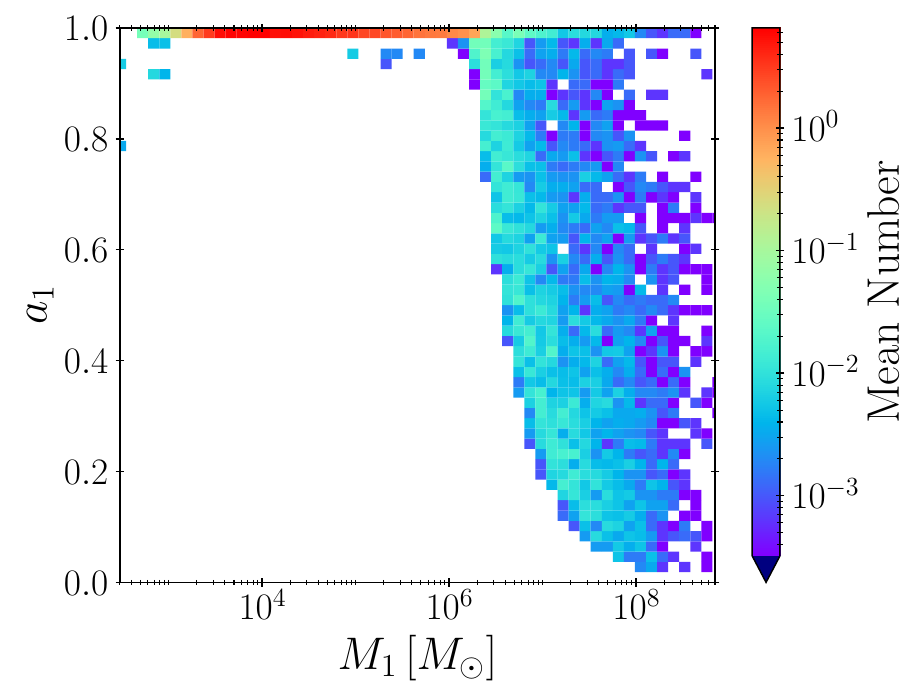}
&
\includegraphics[width=0.4\textwidth]{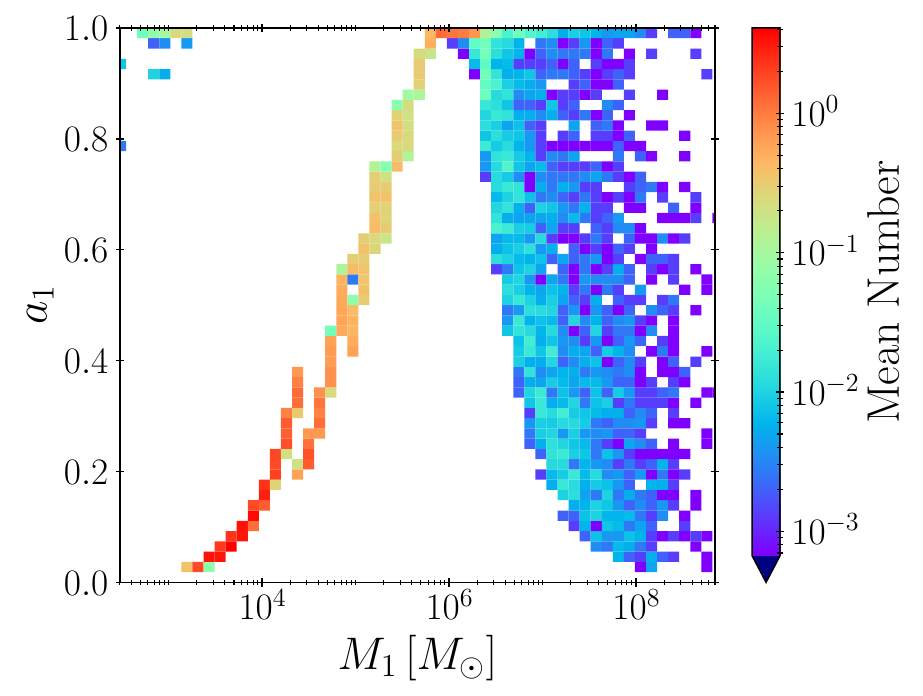}

\end{tabular}

\caption{Left column: Mean number of detectable binaries (SNR $\geq20$) that merge during 4 years in 2D bins of primary mass and spin. Right column: Same as the left, but assuming the existence of a vector boson with mass $\approx 5.1 \times 10^{-16}$ eV and a spin-down timescale of $4.5 \times 10^7$ years. Each panel row corresponds to a black hole population model: Q3delays, Q3nodelays, and PopIII.
Note the x-axis for PopIII extends to lower values than the other models.
}\label{fig.2d_distr}
\end{figure*}

\section{Results}\label{sec.Results}
\begin{figure*}[htbp] 
    \centering
    \includegraphics[width=0.4\textwidth]{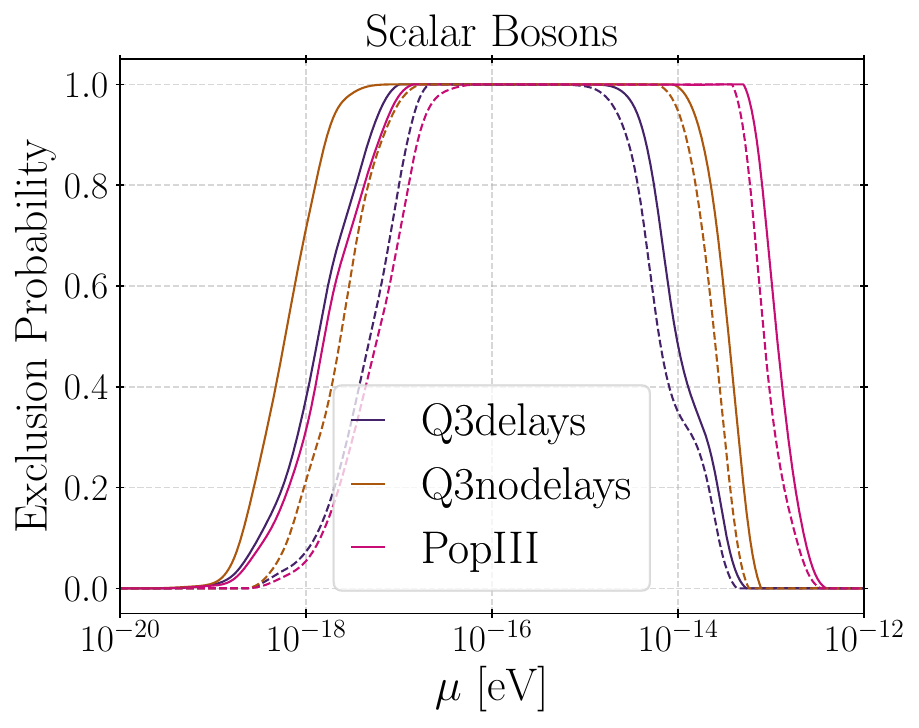}
    \includegraphics[width=0.4\textwidth]{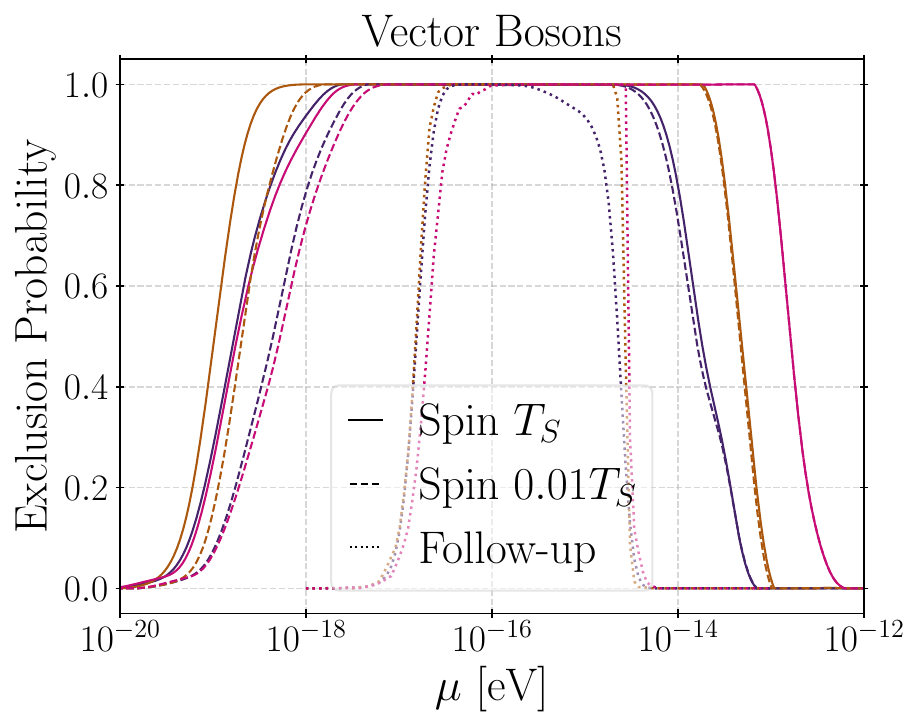}
    \caption{Probability of constraining scalar (left) and vector (right) boson masses from spin measurements of merging binaries in the LISA band. In the vector case (right panel), we also show constraints from follow-up searches for boson-cloud signals. Solid lines correspond to exclusion probabilities from spin measurements assuming an SNR threshold of 20 for the binary, a spin uncertainty of $10\%$ for both binary components and a spin-down timescale  of $4.5 \times10^7$ years. Dashed lines  show the corresponding results for a spin-down timescale of $4.5\times10^5$ years. Dotted lines indicate constraints from follow-up searches, using a SNR threshold of 10. }
    \label{fig.exclusion}
\end{figure*}

\begin{figure*}[htbp] 
    \centering
    \includegraphics[width=0.37\textwidth]{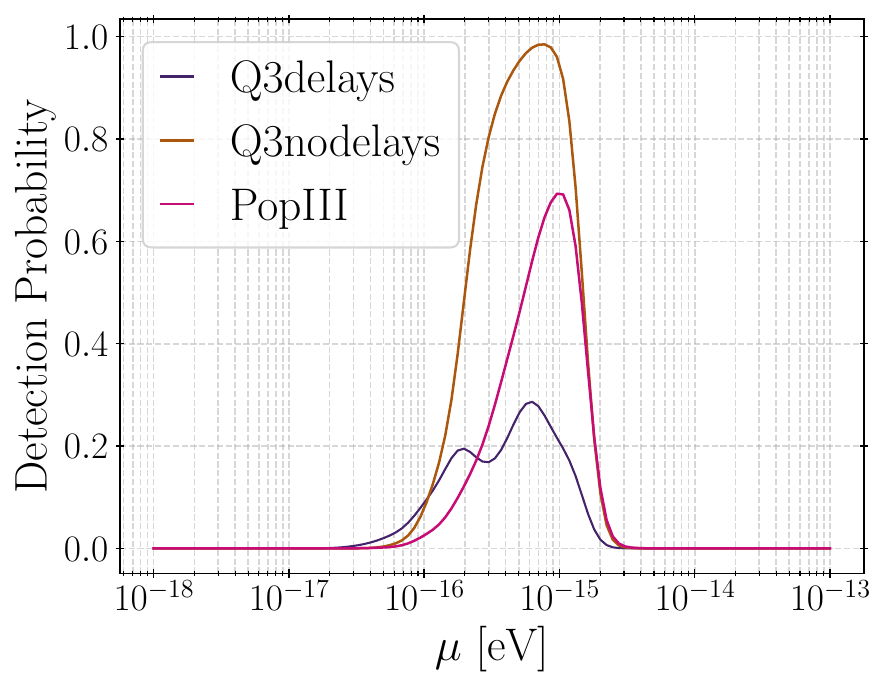}
     \caption{Probability of detecting a gravitational wave signal with SNR $\geq10$ from the dissipation of a vector boson cloud during the four year LISA mission, assuming the existence of such a boson and its impact on both binary components and merger remnant. Results are shown for the three astrophysical population models considered in this work. We assume a spin-down timescale of $4.5 \times 10^7$ years.
     }\label{fig.detection}
\end{figure*}

In this section, we apply the constraint and detection methods described in Sec.~\ref{sec.Method} to the black hole populations introduced in Sec.~\ref{sec.Theory}, and present our main results. We first investigate the range of ultralight boson masses that can be excluded through spin measurements of merging binaries observed by LISA, and through the non-observation of gravitational wave signals from superradiant boson clouds in follow-up searches. The corresponding exclusion regions are shown in Fig.~\ref{fig.exclusion} for both scalar and vector bosons. We then consider the complementary scenario in which an ultralight boson exists, and evaluate the probability of detecting a gravitational wave signal from a superradiant cloud. These results are presented in Fig.~\ref{fig.detection} and apply only to vector bosons.

For all three black hole population models, spin measurements provide stronger constraints than follow-up searches, excluding boson masses over approximately four orders of magnitude. In contrast, follow-up searches are sensitive only to a narrower mass window around $10^{-16}-10^{-15}$ eV. While spin measurements can constrain both scalar and vector bosons, follow-up searches are only effective for vector bosons, since scalar clouds generally grow too slowly to produce observable signals within the LISA mission lifetime. 
Accordingly, the left panel of Fig.~\ref{fig.exclusion} presents the exclusion regions for scalar bosons from spin measurements, while the right panel shows the corresponding constraints for vector bosons from both spin measurements and follow-up searches. 

For the spin-based constraints, we consider two spin-down 
timescales, $\tau_{\rm sd}$: the Salpeter time $T_S= 4.5 \times 10^7$ years, and a more conservative value of $0.01 T_S$. 
The exclusion ranges obtained for $\tau_{\rm sd}=T_S$ are summarized in Table~\ref{tab:spin_constraints}.

The spin-down timescale of $0.01 T_S$ narrows the exclusion region, with the largest reduction at the lower-mass end (although the difference compared to $\tau_{\rm sd}=T_S$ is still quite modest). 
This behavior reflects the fact that lower boson masses correspond to smaller values of $\alpha$, producing more extended boson clouds and longer instability growth times. Therefore, in the case where the spin-down timescale is $0.01 T_S$, some black holes cannot be spun down sufficiently. This effect is particularly pronounced for scalar bosons, whose superradiant instabilities are intrinsically slower than those of vector bosons.
The characteristic black hole masses of the three populations, shown in Fig.~\ref{fig.1d_distr}, determine where this reduction in sensitivity occurs. PopIII peaks around $10^4\,M_\odot$, while Q3delays and Q3nodelays are dominated by black holes with masses of $10^5$-$10^6\ M_\odot$. Since $\alpha = M\mu$, these populations probe different boson mass ranges, leading to the different exclusion limits shown in Fig.~\ref{fig.exclusion}.

A small reduction in the excluded mass range is also observed at the high-mass end, particularly for scalar bosons. In this regime, the dominant superradiant instability shifts to modes with larger azimuthal numbers, whose growth times become progressively longer. As a result, the instability again becomes too slow to efficiently extract black hole spin on the shorter spin-down timescale, reducing the corresponding exclusion region.

\begin{table*}[t]
\centering
\renewcommand{\arraystretch}{1.25}
\caption{Boson mass ranges excluded with probability $>0.99$ from spin measurements for both scalar and vector bosons  assuming a spin-down timescale $\tau_{\rm sd}=T_S$ and from follow-up searches for vector bosons for the three black hole population models.}
\label{tab:spin_constraints}
\begin{tabular}{lccc}
\hline
Population & Scalar via Spin Measurement [eV] & Vector via Spin Measurements [eV] & Vector via Follow-up [eV] \\
\hline
Q3delays   & $[10^{-17},\,2\times10^{-15}]$ & $[2\times10^{-18},\,3\times10^{-15}]$ & $[4\times10^{-17},3\times10^{-16}]$  \\
Q3nodelays & $[5\times10^{-18},\,10^{-14}]$ & $[6\times10^{-19},\,2\times10^{-14}]$ & $[3\times10^{-17},2\times10^{-15}]$ \\
PopIII     & $[3\times10^{-17},\,5\times10^{-14}]$ & $[3\times10^{-18},\,7\times10^{-14}]$ & $[10^{-16},3\times10^{-15}]$ \\
\hline
\end{tabular}
\renewcommand{\arraystretch}{1}
\end{table*}

\begin{figure*}[htbp] 
    \centering
    \includegraphics[width=0.4\textwidth]{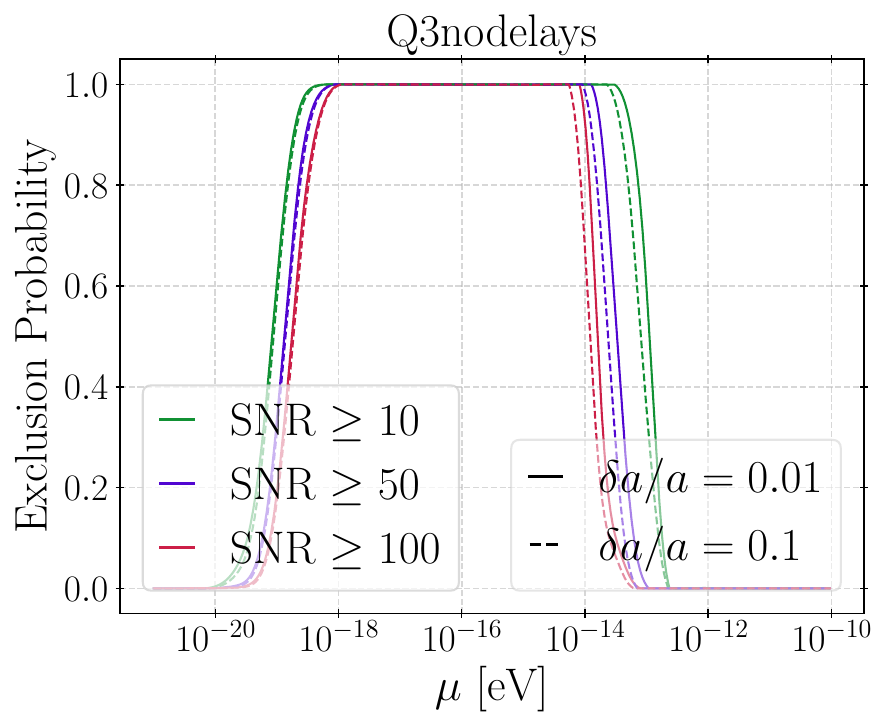}

    \caption{Exclusion curves from spin measurements of both binary constituents for the Q3nodelays population model in the vector case. Results are shown for binary SNR thresholds of 10, 50, and 100 and for spin measurement uncertainties of $10\%$ (solid lines) and $1\%$ (dashed) assuming a spin-down timescale of $4.5 \times 10^7$ years.}  
    \label{fig.spindown}
\end{figure*}

Figure~\ref{fig.spindown} depicts how the constraints from spin measurements depend on the merger SNR threshold and spin measurement accuracy for the Q3nodelays model in the vector boson case. We consider SNR thresholds of $10$, $50$, and $100$, and relative spin uncertainties of  $1\%$ and $10\%$. As expected, higher SNR thresholds lead to narrower constraints, since fewer binaries contribute to the analysis, in particular, around the peak of the detectable merger distribution shown in Fig.~\ref{fig.2d_distr}. 
The high mass end of the exclusion region shows a mild dependence on the assumed spin accuracy. In this high-mass regime, the fastest growing superradiant modes correspond to higher azimuthal numbers $m>1$, whose growth times are longer than those of the $m=1$ mode. As a result, efficient spin extraction requires a longer time, and the corresponding exclusion region becomes more sensitive to uncertainties in the measured spins. However, in general the constrained mass range is much less sensitive to either the SNR or spin accuracy assumption compared to the population model, which justifies our simplistic treatment of gravitational wave spin measurements. 

\begin{figure*}[htbp]
\centering

\begin{tabular}{c c}

\textbf{Follow-up exclusion contribution} & \textbf{Detection contribution} \\[0.3em]

\makebox[0pt][r]{\raisebox{2.2cm}{\rotatebox{90}{Q3delays}}\hspace{0.5em}}%
\includegraphics[width=0.4\textwidth]{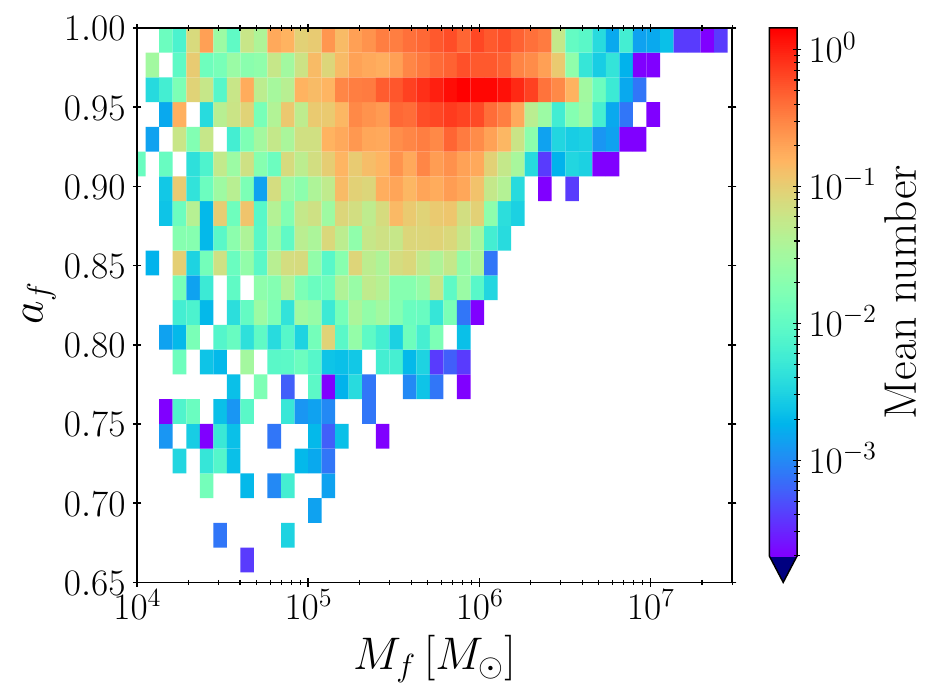}
&
\includegraphics[width=0.4\textwidth]{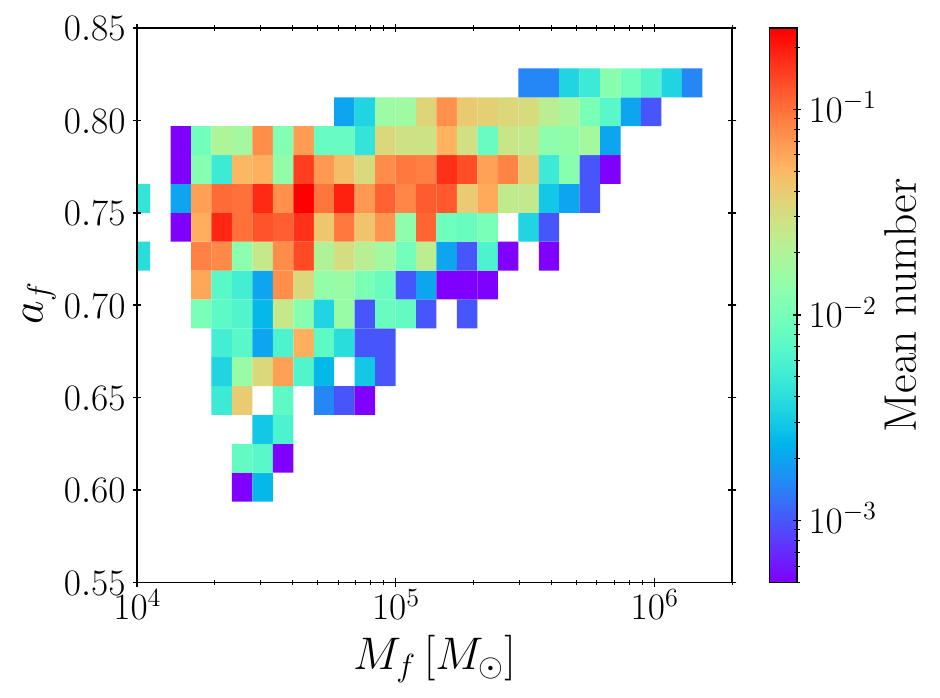}
\\[0.6em]

\makebox[0pt][r]{\raisebox{2.2cm}{\rotatebox{90}{Q3nodelays}}\hspace{0.5em}}%
\includegraphics[width=0.4\textwidth]{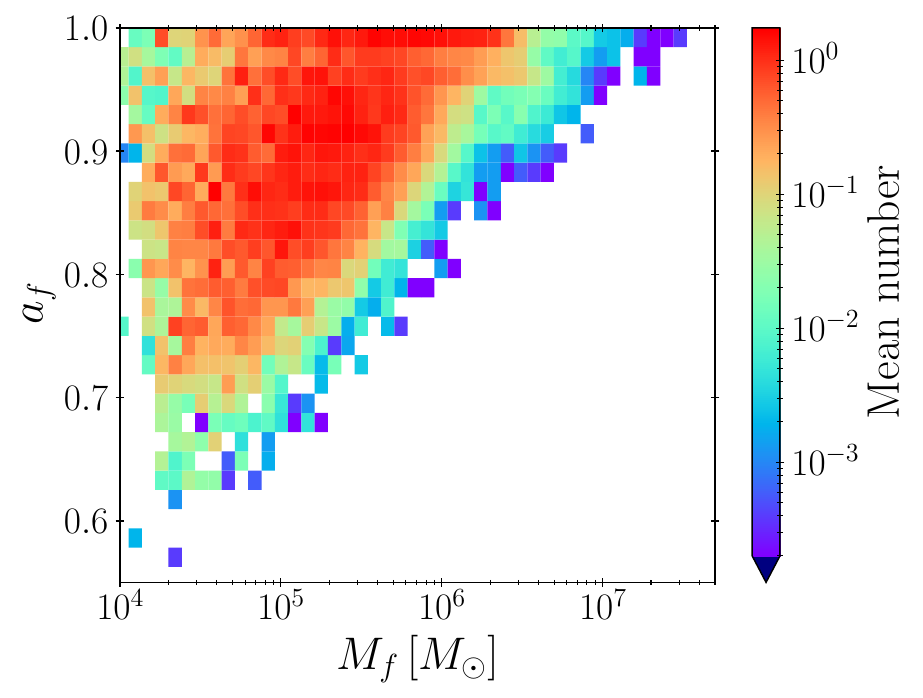}
&
\includegraphics[width=0.4\textwidth]{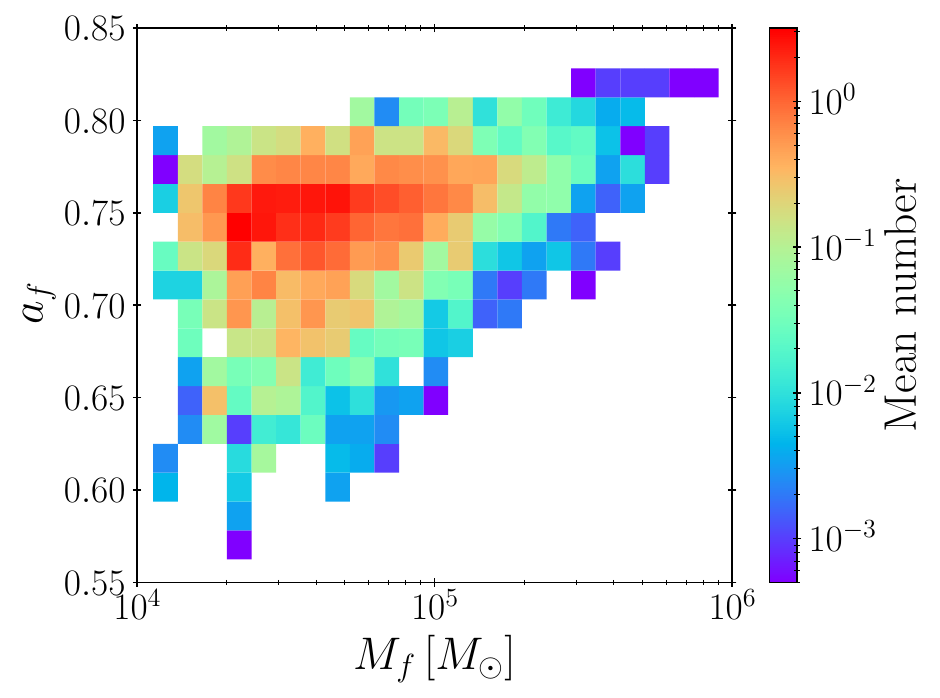}
\\[0.6em]

\makebox[0pt][r]{\raisebox{2.2cm}{\rotatebox{90}{PopIII}}\hspace{0.5em}}%
\includegraphics[width=0.36\textwidth]{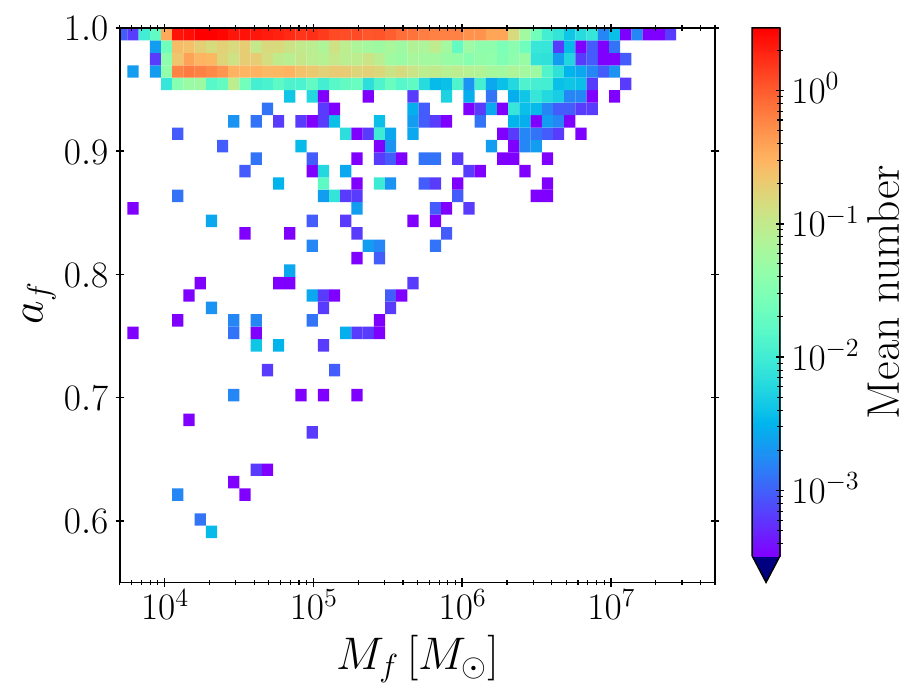}
&
\includegraphics[width=0.4\textwidth]{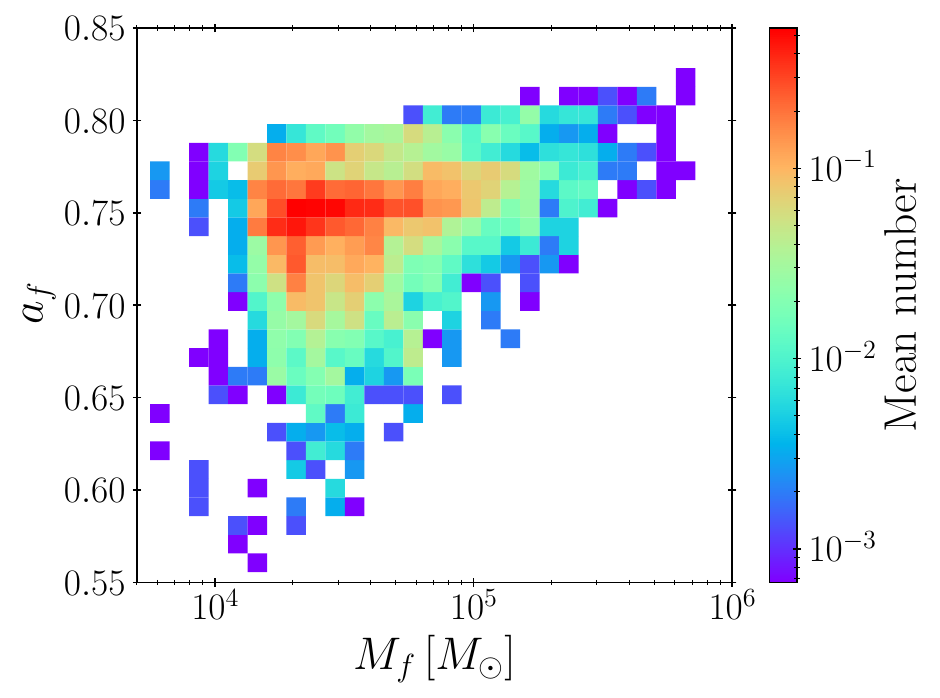}

\end{tabular}

\caption{Mean number of detectable binaries (SNR $\geq20$) merging during the four-year LISA mission as two-dimensional distributions of remnant mass and spin. Left panels show the populations contributing to follow-up exclusion constraints (Fig.~\ref{fig.exclusion}), while right panels show the population contributing to follow-up detection prospects (Fig.~\ref{fig.detection}).  Results are shown for the Q3delays (top), Q3nodelays (middle), and PopIII (bottom) population models.}\label{fig.2d_follow}

\end{figure*}

In contrast, follow-up searches constrain a significantly narrower subset of boson masses compared to spin-based measurements. For probability  $>0.99$, the excluded mass ranges are $[4\times10^{-17},3\times10^{-16}]$ eV for Q3delays, $[3\times10^{-17},2\times10^{-15}]$ eV  for Q3nodelays and $[10^{-16},3\times10^{-15}]$ eV for PopIII. 
As shown in Fig.~\ref{fig.1d_distr}, the remnants contributing to follow-up gravitational wave signals (with SNR $\geq 10$), represent a small subset of all merger remnants. They are typically characterized by masses near the peaks of the population distributions, high remnant spins, and relatively low redshifts. Because the follow-up signals are quasi-monochromatic and generally weaker than the merger signals, only a small fraction falls within the LISA sensitivity band with sufficient SNR. As a result, the exclusion regions are driven by a restricted region of the mass-spin-redshift parameter space, illustrated in the left panel of Fig.~\ref{fig.2d_follow} for the three black hole population models.

The differences between population models can be understood from the underlying distributions in Fig.~\ref{fig.1d_distr} and Fig.~\ref{fig.2d_distr}. The Q3nodelays model exhibits a broad distribution in black hole masses, spins, and redshifts that encompasses that of Q3delays, leading to generally stronger and more extended constraints. By contrast, the PopIII model is dominated by lower mass black holes, and therefore probes higher boson masses, consistent with the inverse scaling between most efficiently probed boson mass and the black hole mass. This trend is also visible in the two-dimensional distributions shown in the left panel of Fig.~\ref{fig.2d_follow}. 

We also consider the scenario in which an ultralight vector boson exists and triggers a superradiant instability affecting the binary constituents as well as the post-merger remnant. In Fig.~\ref{fig.detection}, we present the probability of detecting follow-up signals, assuming a spin-down timescale of $T_S=4.5 \times 10^7$ years. This choice corresponds to a timescale comparable to the rough estimate for accretion timescale. Longer spin-down timescales would allow for more efficient spin extraction by the cloud, resulting in lower remnant spins and weaker signals, and therefore reducing the prospects for detection. Conversely, shorter timescales would preserve higher spins and enhance the detectability prospects. Overall, the detection prospects are limited. In the boson mass range $10^{-16}-10^{-15}$ eV, the detection probability exceeds $\sim 20\%$ for the Q3delays model and $\sim 40\%$ for the Q3nodelays model, reaching $\gtrsim 80\%$ in the range $[2\times10^{-16},1.5\times10^{-15}]$ eV for the latter. For the PopIII model, the detection probability exceeds $\sim 20\%$ reaching $\sim 40\%$ in the range $[5 \times 10^{-16},1.5\times10^{-15}]$ eV. Among the three models, Q3nodelays yields the most optimistic detection prospects. 

A key limitation arises from the fact that the same superradiant instability responsible for generating post-merger signals also acts on the binary components prior to merger. As a result, many systems experience substantial spin-down before coalescence, leading to remnants with significantly reduced spins.
This effect is illustrated in the right panel of Fig.~\ref{fig.2d_distr} for a vector boson with mass $5\times 10^{-16}$ eV. While the black hole mass distribution remains largely unchanged, the spin distribution shifts toward lower values. This reduction in remnant spins has a direct impact on detectability. Lower spins suppress both the growth rate of the superradiant instability and the strength of the resulting gravitational wave signal, making detection more challenging. 
This is particularly important because high spin remnants dominate the detectable follow-up population, as shown in Fig.~\ref{fig.2d_follow}. The left panel indicates that remnants contributing to follow-up exclusions typically have spins $\gtrsim 0.9$ for PopIII and Q3delays, while Q3nodelays also includes a broader range extending to lower spins. The right panel shows that detectable signals arise from a narrower spin range, typically $0.7$-$0.8$, slightly above the canonical value $\simeq 0.7$ expected for the merger of an equal-mass, non-spinning binary.

In summary, LISA observations offer strong prospects for constraining ultralight scalar and vector bosons through black hole spin measurements, with sensitivity spanning several orders of magnitude in boson mass. In contrast, direct detection, through post-merger follow-up searches is considerably more challenging. Scalar clouds evolve too slowly to produce observable signals, while in the vector case, the pre-merger spin-down of the progenitor black holes suppresses the population of high-spin remnants required for detectable gravitational wave signals.
Nevertheless, ultralight bosons may still leave observable indirect signatures, such as characteristic features in the black hole mass-spin distribution, as illustrated in the right panel of Fig.~\ref{fig.2d_distr}. This effect is most pronounced in scenarios such as PopIII and Q3delays, which predict a larger fraction of rapidly spinning systems susceptible to superradiant spin-down.


\section{Discussion and Conclusion}\label{sec.Discussion}
In this work, we have investigated the potential to probe ultralight bosons through black hole superradiance using observations of massive black hole binary mergers with LISA. We have considered spin measurements of merging binaries, as well as follow-up searches for gravitational wave emission from superradiant clouds around merger remnants, to estimate both exclusion regions and detection prospects. 

We find that across the three astrophysical models for massive black hole population, namely Q3delays, Q3nodelays, and PopIII, introduced in Ref.~\cite{Barausse:2012fy} and updated in Ref.~\cite{Barausse_2023}, spin measurements alone can constrain ultralight boson masses over approximately four orders of magnitude. For scalar bosons, the constrained range is approximately $[5\times10^{-18},10^{-14}]$ eV while for vector bosons it extends down to approximately $[6\times10^{-19},2\times10^{-14}]$ eV. These ranges vary by up to almost an order of magnitude depending on the underlying black hole population model and the assumed spin-down timescale. In addition, the absence of signals in follow-up searches provides complementary constraints for vector bosons, particularly around $3\times10^{-17}-3\times10^{-15}$ eV. Follow-up searches are not effective for scalar bosons, as their cloud growth timescales typically exceed the expected LISA mission duration. While the exclusion probabilities are consistently high across all models, the precise mass ranges vary due to differences in the distributions of black hole masses, spins and redshifts. In contrast, detection probabilities are significantly more model dependent, reflecting the sensitivity of follow-up signal detectability on the detailed population properties. Given current uncertainties in the massive black hole population, these results should be interpreted as model dependent forecasts. 

Our forecasts on constraints on scalar bosons through black hole spin measurements are in good agreement with those in Refs.~\cite{Brito:2017zvb,Brito:2017wnc}. The latter report expected constraints in the mass range $[4\times10^{-18},10^{-14}]$ eV and $[10^{-18},2\times10^{-13}]$ eV respectively. We should note that they are using the original catalogs \cite{Barausse:2012fy,Klein:2015hvg} for massive black hole populations, while we use more recently updated ones. They also explore the constraints that can be put through blind coherent and stochastic  searches of the quasi-monochromatic gravitational waves from scalar boson cloud dissipation (while the only additional observations we consider are targeted follow-up searches). It would be interesting to consider the prospects for blind searches for vector boson gravitational wave signals for LISA in the future.
Studies such as Refs.~\cite{Brito:2017zvb,Brito:2017wnc,Ng2021} consider how, by using gravitational wave observations to simultaneously constrain the underlying black hole population model and a possible ultralight boson, one can gather evidence for the existence of an ultralight boson solely from spin constraints.
Such a strategy could also be applied to the case considered here. 

In this work we have adopted a conservative detectability criterion based on the probability of obtaining at least one exclusion event per realization. While this provides a robust and easily interpretable measure of LISA’s constraining power, it does not fully exploit the information contained in multiple constraining binaries within a single realization.
A more refined analysis could incorporate the full distribution of exclusion events per realization, or weight constraints by their measurement uncertainties. In addition, a Bayesian framework combining information from multiple sources could provide a more detailed characterization of the posterior constraints on ultralight boson parameters. In addition, we used a match-filtered SNR to assess the detectability of the post-merger gravitational wave signal. However, the optimal search strategies for such signals are currently uncertain, and further work is required to determine which methods will be realistically feasible for LISA data analysis.

Throughout this work, we have assumed purely gravitational interactions, neglecting both self-interactions of the bosonic fields and couplings to other sectors. For scalar fields, self-interactions of axion-like particles can modify the evolution of superradiant clouds, as discussed in Refs. \cite{Baryakhtar_2021,witte2025steppingsuperradianceconstraintsaxions,Arvanitaki:2010sy,Omiya:2022mwv,Collaviti:2024mvh}. 
These effects can include mode mixing, energy dissipation through bosonic radiation, and nonlinear phenomena such as bosenova collapse.
Depending on the coupling strength, self-interactions may have a negligible impact
or instead halt the exponential growth of the boson cloud at much smaller amplitudes than through purely gravitational interactions.

For vector bosons, one example interaction that has been studied is the addition of a kinetic term between the dark photon and the Standard Model photon \cite{Okun:1982xi,Holdom1986,Feng:2009mn,Cyr-Racine:2012tfp,Bhoonah:2018gjb}. For sufficiently large mixing strength, the dark photon cloud can generate a plasma of pair particles that emit electromagnetic radiation, slowing down the cloud's growth \cite{Siemonsen_2023,xin2024darkmagnetohydrodynamicsblackhole}. Another interaction that has been proposed is that of the dark photon with a massive Higgs-like complex scalar. In this case, the cloud can lose energy through the emission of bosonic radiation \cite{Fukuda_2020} or through the formation of strings that can lead to the explosive disruption of the cloud if the interactions are strong enough \cite{East_2022,East_20222}. 

An important direction for future work is to include such interaction terms and extend the analysis to the joint mass-coupling plane. This would allow for a more comprehensive assessment of how interaction strengths impact both exclusion limits and detection prospects with LISA. One can obtain preliminary estimates, using methods similar to those of Ref. \cite{aswathi2025}. 
Using the sources of the catalogs, we estimate that a quartic self-interaction coupling with $f \gtrsim \mathcal{O}(10^{15}-10^{16})$ GeV would not affect the purely gravitational interaction analysis taken here. Similarly, the spin constraint forecasts would apply to a dark photon-photon kinetic mixing with $\epsilon<\mathcal{O}(0.001)$ or a Higgs-Abelian type of interaction
with $g \lambda^{-1/4}\lesssim\mathcal{O}(10^{-23})$.
These are rough estimates of the constraints on the upper boundaries of these interaction couplings. Most of these interaction strengths are computed assuming small $\alpha$ values and assume only the first azimuthal number is present. A dedicated analysis is required to make these estimates robust. However, it is worth mentioning that for supermassive black holes interactions are less important than for stellar mass black hole systems. Although the occupation number in the former case is larger, the spatial extent of the cloud is also much larger, making it less dense and decreasing the interaction rate. 

Overall, our results highlight the strong potential of LISA to probe ultralight bosons through black hole superradiance, particularly via spin measurements. Direct detection through follow-up searches is also possible, though it appears more challenging, especially once the backreaction of superradiance on the spins of the binary constituents is taken into account. The combined observational strategies provide a powerful and complementary approach to exploring new physics beyond the Standard Model.

\acknowledgements
It is a pleasure to thank Nils Siemonsen for valuable discussions. 
I.G. and W.E. acknowledge support from a Natural Sciences and Engineering Research
Council of Canada Discovery Grant and
an Ontario Ministry of Colleges and Universities Early Researcher Award.
This research was
supported in part by Perimeter Institute for Theoretical Physics. Research at
Perimeter Institute is supported in part by the Government of Canada through
the Department of Innovation, Science and Economic Development and by the
Province of Ontario through the Ministry of Colleges and Universities.
Computations were performed on the Symmetry cluster at Perimeter Institute. 
\newpage

\appendix
\section{Follow-up SNR calculation}\label{app:SNR}
To compute the SNR for the follow-up signals we make use of the \texttt{SuperRad} package \cite{Siemonsen:2022yyf,May:2024npn}.
First we write the two polarizations of the strain in the source frame at a specified luminosity distance using the two most dominant modes in an expansion of the waveform in spin-weighted -2 spherical harmonics:
\begin{align}\label{eq.strain_sf}
h_{+}(t) &\simeq |h_{2,2}(t) ( {}_{-2}S_{2,2}(\theta) + {}_{-2}S_{2,-2}(\theta)) \nonumber \\
&+ h_{3,2}(t) ( {}_{-2}S_{3,2}(\theta)
- {}_{-2}S_{3,-2}(\theta))  |\cos\!\big(\phi(t)\big)
\\
h_{\times}(t) &\simeq |h_{2,2}(t) ( {}_{-2}S_{2,2}(\theta) - {}_{-2}S_{2,-2}(\theta)) \nonumber \\
&+ h_{3,2}(t) ( {}_{-2}S_{3,2}(\theta)
+ {}_{-2}S_{3,-2}(\theta))  |\sin\!\big(\phi(t) + \delta\big)
\end{align}
where,
\begin{equation}
    h_{\ell,m} \simeq \mathcal{A}_{\ell,m} \frac{1}{1 + t/\tau_{\rm GW}}
\end{equation}
and we calculate the right-hand-side quantities with \texttt{SuperRad}. We also approximate the phase in~\eqref{eq.strain_sf} as:
\begin{align}\label{eq.strain_phase}
    \phi(t) &\simeq 2 \pi \int \Big(f_{\infty} + \frac{\delta f}{1 + t/\tau_{\rm GW}}\Big)dt \nonumber \\
    &= 2\pi f_{\infty} t + \delta f \tau_{\rm GW} \ln(1 + t/\tau_{\rm GW})
\end{align}
where $f_{\infty}$ is the GW frequency as the boson cloud mass approaches zero at late times, and $\delta f = f_{\rm GW}(t=0) - f_{\infty}<0$ is the negative frequency shift when the boson cloud mass is maximum. We obtain both quantities from \texttt{SuperRad}. We neglect the phase difference $\delta$ between the two polarizations since we are interested in the magnitude squared of the strain and a constant phase will not matter.

We need to transform the signal to the frequency domain in order to compute the SNR. In order to do so, we will derive the Fourier Transform (FT) of~\eqref{eq.strain_sf} from the FT of the auxiliary function:
\begin{equation}\label{eq.aux_td_sf}
    H(t) = \frac{\mathcal{A}(\theta)}{1 + t/\tau_{\rm GW}}e^{i \phi(t)} \, .
\end{equation}
Since the strain is dimensionless,
the strain in the detector frame is just the strain in the source frame evaluated at the redshifted time $t_d=(1+z)t_s$. 
So the detector frame FT is related to the source frame FT $H(f)$ by
\begin{align}\label{eq.FT}
    \tilde{H}_d(f_d) = (1+z) \tilde{H}(f=(1+z)f_d) \ .
\end{align}
Now we can compute the FT in the source frame and then relate it to the detector frame. From~\eqref{eq.aux_td_sf}, assuming all quantities are in the source frame, we get
\begin{align}\label{eq.FT_sf_1}
    \tilde{H}(f) = \int_{-\infty}^{\infty} dt \frac{\mathcal{A}(\theta)}{1 + t/\tau_{\rm GW}}e^{2 i \pi [ f_{\infty} t + \delta f \tau_{\rm GW} \ln(1 + t/\tau_{\rm GW}) - f t]} .\nonumber \\
\end{align}
We make use of stationary phase approximation following Ref.~\cite{Chan:2022dkt}. The phase in~\eqref{eq.FT_sf_1} $\Phi(t)$ 
will be approximated by an expansion around the stationary point $\dot \Phi(t_0) =0$ where
\begin{equation}\label{eq.SPA_t0}
t_0 = \tau_{\rm GW} \Big( \frac{\delta f}{f - f_\infty} -1 \Big)
\end{equation}
where $f(t)$, the frequency as a function of time, is well defined since the frequency is monotonically increasing with time.
Using the second derivative of the phase at the stationary point,
\begin{equation}
    \ddot\Phi (t_0) = -2 \pi \frac{(f-f_\infty )^2}{\delta f \tau_{\rm GW}} \ , 
\end{equation}
\eqref{eq.FT_sf_1} is approximated by,
\begin{align}\label{eq.FT_sf_2}
    \tilde{H}(f) \simeq\frac{\mathcal{A}(\theta)}{1 + t_0/\tau_{\rm GW}} e^{i  \Phi(t_0(f)) } \int_{-\infty}^{\infty} dt\,  e^{i \frac{1}{2}\ddot\Phi(t_0)(t-t_0)^2 } \, ,
\end{align}
where the last integral is a Fresnel-type integral $\int_{-\infty}^{\infty} dx e^{i \gamma x^2} = \sqrt{\frac{\pi}{\gamma}} e^{i\pi/4}$ for $\gamma>0$. Hence, ~\eqref{eq.FT_sf_2} becomes:
\begin{equation}
    \tilde{H}(f) \simeq \mathcal{A}(\theta) e^{i  \Phi(t_0(f)) } e^{i\pi/4} \sqrt{ -\frac{\tau_{\rm GW}}{\delta f}} \, .
\end{equation}
The Fourier transforms of $h_+(t)$ and $h_\times(t)$ are related to that of $H(t)$ via
\begin{align}
    \tilde{h}_+(f) &= \frac{\tilde{H}_+(f) + \tilde{H}_+^*(-f)}{2} \nonumber \\
    \tilde{h}_\times(f) &= \frac{\tilde{H}_\times(f) - \tilde{H}_\times^*(-f)}{2i}
    .
\end{align}
We discard the negative frequency part since the SNR is an integral over positive frequencies and has a factor of 2 in front of it accounting for the negative frequency plane, and while being careful in adjusting $\mathcal{A}(\theta)$ for each polarization. Now we can construct the magnitude squared that is required to compute the SNR as defined in \eqref{SNR}, which will be equal to $|\tilde{h}_+(f)|^2 + |\tilde{h}_\times(f)|^2$ since they are orthogonal. We then want to take the inclination average:
\begin{align}\label{eq.inclination_averaging}
    \left<|\tilde{h}_{+_d}(f_d)|^2\right>_{\theta}
    &= \frac{1}{2} \int_0^\pi
    \tilde{h}_{+_d}(f_d)\tilde{h}_{+_d}^*(f_d)
    \sin\theta\, d\theta
    \nonumber \\
    \left<|\tilde{h}_{\times_d}(f_d)|^2\right>_{\theta}
    &= \frac{1}{2} \int_0^\pi
    \tilde{h}_{\times_d}(f_d)\tilde{h}_{\times_d}^*(f_d)
    \sin\theta\, d\theta
\end{align}
where we have to be careful of the two modes used and the cross-terms for each polarization expression. 

For the binary SNR, we use \texttt{lisabeta} ~\cite{Marsat:2020rtl} to compute the frequency domain strain amplitude $\tilde h_b (f_{d_b})$ of the $\ell=m=2$ mode in the detector frame. The contribution of the $\ell=2,m=-2$ mode in the SNR calculation is identical to the one from $\ell=m=2$, since for non-precessing, quasi-circular binaries their magnitude is the same. The frequency domain strain amplitude does not contain any phase, hence is sky location and polarization independent. We then also integrate it over inclinations as in Eq.~\eqref{eq.inclination_averaging} and compute the SNR using Eq.~\eqref{SNR}.

\bibliographystyle{apsrev}  
\bibliography{references}

\end{document}